\documentclass[onecolumn,amsmath,amssymb,aps]{revtex4-2}

\usepackage{graphicx}
\usepackage{dcolumn}
\usepackage{bm}
\usepackage[utf8x]{inputenc}
\usepackage{float}
\usepackage{bbold}

\begin{document}


\title{The Status of $\Xi_{cc}^{++}$ Baryon: investigating quark-diquark model}

\author{Halil Mutuk}
 \email{hmutuk@omu.edu.tr}
 \affiliation{Department of Physics, Faculty of Arts and Sciences, Ondokuz Mayis University, 55139, Samsun, Turkey}

\begin{abstract}
Inspired by the discovery of the doubly-charmed and doubly-charged $\Xi_{cc}^{++}$ baryon, in this present work, we investigate spin-1/2 and spin-3/2 mass spectra and magnetic moments of $ccu$ baryon in quark-diquark model with a nonrelativistic potential. In the quark-diquark picture, the three-body problem of the $ccu$ baryon is reduced to the two-body bound state problem. We obtained ground and radially excited mass spectra and magnetic moments of the ground state. Ground state mass spectra agree well with the results in the literature, especially with lattice QCD computations. Predictions for the magnetic moment are compatible with the literature. According to our results, $cu$ diquark clustering in the $\Xi_{cc}^{++}$ baryon is more favorable than the $cc$ diquark clustering.
\end{abstract}

\pacs{Later}
\maketitle

\section{Introduction}
Recently, significant experimental studies have been conducted in heavy baryon spectroscopy, which yielded many observations. Many experimental collaborations such as CLEO, Belle, BABAR, LHCb, CDF, D0, CMS are devoted to studying baryons with heavy quarks, $c$ and $b$. \cite{Artuso:2000xy,Mizuk:2004yu,Acosta:2005mq,Aaltonen:2007ap,Abazov:2007am,Aubert:2007dt,Aubert:2007bt,Aaltonen:2007ar,Abazov:2008qm,Aaltonen:2009ny,Abazov:2011wt,Aaltonen:2011sf,Chatrchyan:2012ni,Aaij:2012da,Aaij:2016wxd}. 

Baryons are described as bound states of three constituent quarks. These are effective degrees of freedom which mimic the three valence quarks inside baryons, with virtual gluons and quarkonium $q\bar{q}$ sea pairs. Baryons $(Qqq)$ made of light $(q)$ and heavy $(Q)$ flavor quarks  are important topics because of the existence of various mass scales of quarks. Theoretical and experimental  investigations for properties of singly-heavy baryons can pave the way for doubly and triply heavy baryons. Singly-heavy baryons  have been discovered, and quantum numbers of most of the observed states have been assigned \cite{Zyla:2020zbs}. 

In 2017, the LHCb Collaboration reported the first observation of the doubly-charmed baryon $\Xi_{cc}^{++}$ with a mass around 3621 MeV \cite{Aaij:2017ueg,Aaij:2018gfl}. It has a possible quark content of $ccu$, and it was observed in two exclusive decay modes; $\Xi_{cc}^{++}\rightarrow \Lambda_c^+ K^- \pi^{+} \pi^{+}$ and $\Xi_{cc}^{++}\rightarrow \Xi_{c}^{+}\pi^{+}$. This doubly-charmed baryon signal was ``observed" ten years ago by the SELEX Collaboration in the $\Lambda_c^+ K^- \pi^{+}$ final decay \cite{Mattson:2002vu,Ocherashvili:2004hi}. The results include a number of unexpected features, such as a notable short lifetime and a large production rate relative to that of  $\Lambda_c^+$ baryon. However, FOCUS \cite{Ratti:2003ez}, BABAR \cite{Aubert:2006qw} and BELLE \cite{Chistov:2006zj} Collaborations did not confirm  this state with the properties reported by SELEX Collaboration at that time. Prior to the LHCb observation, the existence of $\Xi_{cc}^{++}$ baryon was unclear. 

The observation of $\Xi_{cc}^{++}$ baryon was a milestone in hadron physics theoretically and experimentally. Before this observation, many studies about doubly-charmed baryons have been conducted in the literature \cite{Bagan:1992za,Roncaglia:1995az,Ebert:1996ec,Tong:1999qs,AliKhan:1999yb,Itoh:2000um,Gershtein:2000nx,Kiselev:2001fw,Narodetskii:2001bq,Mathur:2002ce,Lewis:2001iz,Ebert:2002ig,Flynn:2003vz,JuliaDiaz:2004vh,Chiu:2005zc,Brambilla:2005yk,Migura:2006ep,Faessler:2006ft,Albertus:2006ya,Liu:2007fg,Roberts:2007ni,Valcarce:2008dr,Liu:2009jc,Namekawa:2012mp,Alexandrou:2012xk, Aliev:2012ru,Aliev:2012iv,Namekawa:2013vu,Can:2013tna,Sun:2014aya,Chen:2015kpa,Sun:2016wzh,Shah:2016vmd,Chen:2016spr,Shah:2017liu,Kiselev:2017eic}. This observation also triggered many new studies \cite{Chen:2017sbg,Yao:2018ifh,Ozdem:2018uue,Aaij:2018wzf,Bahtiyar:2018vub,Meng:2018zbl,Wang:2019mhm,Aaij:2019dsx,Gutsche:2019iac,Gutierrez-Guerrero:2019uwa,Li:2020uok,Shi:2021kmm,Li:2021rfj,Aliev:2020lly}.

Doubly heavy baryons $(QQq)$, where $Q=c,b$, and $q=u,d,s$, represent a unique aspect of baryons since they contain one light quark and two heavy quarks. The doubly-heavy ($QQq)$ baryon should be described in terms of a combination of perturbative and non-perturbative QCD due to the substantially different energy scales of the masses of heavy and light  quarks. The presence of one light quark may provide a field to study the dynamics of the light quark in the vicinity of heavy quarks in the baryon. Heavy baryon investigations are important for understanding the basic properties of QCD, such as flavor independence, the short-range of interquark potential, the confining mechanism, three body forces inside the bound state, and relevant fundamental degrees of freedom.  In this perspective, the $ccu$ baryon may present information on quark dynamics with the superposition of the slow motion of two heavy quarks experiencing the short-range QCD potential together with a fast-moving light quark surrounding them \cite{Fleck:1989mb}. In other words, the dynamics of light quark in the vicinity of two heavy quarks inside the baryon may provide a laboratory to understand the nature of heavy quark physics. 

Quark models explain several properties of baryons quite well, such as mass spectra, strong decays, magnetic moments, and form factors. Notwithstanding, quark model predicts many more states than the experimentally observed states. This phenomenon is known as the \textit{missing resonance problem}. An alternative way to overcome this missing resonance problem is to consider models that are characterized by a smaller number of effective degrees of freedom concerning the three-quark model for baryon. This is the inspiration for the idea of the quark-diquark model for baryon.

Heavy-quark baryons are ideal systems to probe QCD dynamics  and the general structure of hadronic systems. Baryons are described by calculating masses, energies, magnetic moments, decay widths, and etc. Among these properties, magnetic moments give information about the internal structures, which provides valuable information regarding the inner structures of hadrons and helps to understand the mechanism of strong interactions at low-energy.

Since $\Xi_{cc}^{++}$ baryon comprises two charm quarks, it can be studied using the nonrelativistic quark model. In the nonrelativistic limit, QCD interaction can be reduced to the potential interaction between quarks. In general, the potential interaction includes linear confinement and Coulomb-type terms, and higher-order
terms such as the spin-spin  interaction, tensor interaction, and spin-orbit interactions. 

In this present paper, we study mass spectra of ground and excited states, magnetic moments, and diquark clustering of the $ccu$ baryon with spin-1/2 and 3/2. The paper is organized as follows: in Section \ref{sec:level2}, we describe the model of this work. In Section \ref{sec:level3}, we present our results of mass spectra, magnetic moments and baryon-mass inequalities. Section \ref{sec:level4} is reserved for concluding remarks.

\section{\label{sec:level2}Methodology}
\subsection{Quark-Diquark Model}
The idea of diquarks is almost as old as quarks. The possibility of pairwise clustering of quarks in baryons has been studied since the early days of the quark model \cite{Ida:1966ev,Lichtenberg:1967zz,Lichtenberg:1981pp}. See \cite{Santopinto:2004hw,Ferretti:2011zz,Santopinto:2014opa,DeSanctis:2014ria} for recent studies.  A diquark is defined as a colored bound state of two quarks, although it has not been observed yet. However, the lack of experimental observation does not disprove the hypothesis of diquarks as constitutive building blocks inside the baryons. 

There are some indications that the quark-diquark approximation is plausible for a doubly-heavy $(qQQ)$ baryon. The first indicator is related to mean separations. The dynamics in the $(qQQ)$ baryon require the mean separation of the $(QQ)$ pair to be significantly smaller than the mean distance of the $q$ from the center of mass of the $(QQ)$ pair. The second one is that there is no need to antisymmetrize the wave function under the interchange of the coordinates of heavy and light quarks \cite{Anselmino:1992vg}. Thirdly, assuming that the baryon is composed of a quark and diquark  reduces the degrees of freedom inside the baryon. If one considers diquark as antiquark inside a baryon, a three-body problem turns into a two-body problem that simplifies mathematical calculations. This is not at all a mathematical simplification. Internal dynamics among quark-diquark constituents can be described by a single relative coordinate, say $\textbf{r}$, instead of the usual $\rho$ and $\lambda$ Jacobi coordinates of a three-quark system \cite{Barabanov:2020jvn}. It was mentioned in Ref. \cite{Richard:1992uk} that quark-diquark approximation is suited for angular excitations of ordinary baryons. Indeed in the framework of the nonrelativistic quark model, diquark clustering in baryons was studied \cite{Fleck:1988vm}. They found that diquark structure emerges clearly in $(qQQ)$ baryons with low angular momentum. 

As a result of group theory, two quarks can attract one another in the $\bar{3}$ representation of SU(3) color; thus, a diquark forms having the same color features of an antiquark. In other words, a quark inside a baryon sees a color $\bar{3}$ set of two quarks which is analogous to the antiquark seen by a quark in a conventional meson. 
This suggests that the interaction between a quark and a diquark in a baryon can be studied similarly to the one between a quark and an antiquark in a meson. 

The $cc$ diquark is a heavy system and relative slow motion of the tightly bound color antitriplet $cc$ diquark in $ccu$ baryon is similar to quarkonium $(Q\bar{Q})$ since heavy diquark acts as a static color source for the third constituent light quark $q$. The $(cc)-(u)$ structure may be described similarly to $\bar{Q}q$ mesons, where the $cc$ pair plays the role of the heavy antiquark. Indeed it can be expected that two heavy quarks in the diquark are very close to each other so that they are seen as a whole structure by the light quark.

In the quark-diquark model, two heavy quarks inside a baryon are thought to form a heavy diquark acting as a static color source for the third constituent light quark. Heavy-quark current masses are clear signals of chiral symmetry breaking. As a result of this phenomenon, no Goldstone-boson exchange occurs, and the perturbative short-range one-gluon-exchange (OGE) and confinement dominate the interaction potential. In doubly-heavy (doubly-charmed in our case) baryons, chiral symmetry is explicitly broken. As a consequence, the interaction potential reduces to the OGE exchange and confinement. In an early work, masses of ground state spin-3/2 baryons were calculated in a relativistic quark-diquark model with a potential motivated by QCD \cite{Lichtenberg:1982jp}. The so-called Cornell potential meets these properties. Such a choice with a different potential model was made in \cite{Vijande:2004at}. They predicted the $ccq$ and $ccc$ spectra employing a potential model that describes the light-baryon and the heavy and heavy-light meson spectra. In \cite{Giannuzzi:2009gh} the spectra of baryons having two heavy quarks are obtained in  light quark-heavy diquark model with a semirelativistic quark model by using a potential obtained in a gauge-gravity under the assumption of one-gluon exchange. 

In order to study $\Xi_{cc}^{++}$  baryon, we use Cornell potential.  The Cornell potential has been giving reliable and precise results in describing hadronic phenomenology. It reads as
\begin{equation}
V(r)=\frac{\kappa \alpha_s}{r} + br, \label{Cornell}
\end{equation}
where $\kappa$ is a color factor and is associated with the color structure of the system, $\alpha_s$ is the fine-structure constant of QCD, and $b$ is the string tension. The first term (Coulomb term) in Eq. (\ref{Cornell}) arises from the OGE. It is associated with a Lorentz vector structure. The second term (linear part) is responsible for confinement, which is usually associated with a Lorentz scalar structure.

In the center-of-mass frame, using spherical coordinates, one can factorize the angular and radial parts of the Schrödinger equation. Let $\mu\equiv m_1m_2/(m_1+m_2)$, where $m_1$ and $m_2$ are the constituent masses of quark 1 and quark 2, respectively. Then radial Schrödinger equation can be written as
\begin{equation}
    \left\{  \frac{1}{2\mu}\left[-\frac{\text{d}^2}{\text{d}r^2}+\frac{L(L+1)}{r^2}\right] +V(r) \right\} \psi(r) = E\psi(r).
    \label{schtrans}
\end{equation}
A spin-spin interaction can be included under the assumption of the Breit-Fermi Hamiltonian for OGE \cite{Lucha:1991vn} as:
\begin{align}
    V_S(r) &= -\frac{2}{3(2\mu)^2}\nabla^2V_V(r)\mathbf{S}_1\cdot\mathbf{S}_2 \nonumber\\
    &= -\frac{2\pi\kappa\alpha_S}{3\mu^2}\delta^3(r)\mathbf{S}_1\cdot\mathbf{S}_2 \,.
    \label{spinspin}
\end{align}
The spin-spin interaction in the zeroth-order (unperturbed) potential can be written by replacing the Dirac delta with a Gaussian function 
\begin{equation}
    V_S(r) = -\frac{2\pi\kappa\alpha_S}{3\mu^2}\left(\frac{\sigma}{\sqrt{\pi}}\right)^3\exp\left(-\sigma^2 r^2\right)\mathbf{S}_1\cdot\mathbf{S}_2,
    \label{spinsmear}
\end{equation}
which introduces a new parameter $\sigma$.

With this new definition, the Schrödinger equation of the form of Eq.~(\ref{schtrans}) can be written in a compact form as follows:
\begin{equation}
    \left[  -\frac{\text{d}^2}{\text{d}r^2} +V_{\text{eff}}(r) \right]\varphi(r) = 2\mu E\varphi(r).
    \label{schtranss}
\end{equation}
Here the effective potential $V_{\text{eff}}(r)$ is given by
\begin{equation}
    V_{\text{eff}}(r)\equiv 2\mu\left[ V(r)+V_S(r) \right] + \frac{L(L+1)}{r^2}
    \label{veff}
\end{equation}
and composed of Cornell potential, spin-spin interaction and orbital excitation. 

Computing baryon masses in the quark-diquark model has two steps. In the first step, the diquark mass is calculated. In the second step, the mass of the system formed by a diquark and a quark is calculated. As mentioned before, the interaction of a quark-diquark pair is assumed to be the same as that of a quark-antiquark pair. The nonrelativistic Schrödinger equation with Cornell potential is solved and the mass spectrum is obtained using
\begin{equation}
M_B=m_{QQ} + m_q + \left\langle H \right\rangle,
\end{equation}
where $m_{QQ}$ is the mass of heavy diquark, $m_q$ is the mass of quark and $\left\langle H \right\rangle$ is composed of Cornell potential and the spin dependent interactions.

\subsection{Magnetic Moment}
Electromagnetic properties correspond to the major and meaningful parameters of the doubly-heavy baryons. Magnetic moment is one of the most studied properties of the  baryons since it provides information about the internal structures and shape deformations. These properties give valuable information about describing inner structures of hadrons and strong interaction mechanism at low energies. Specifying the magnetic moment provides important information on structure, size, and shape of hadrons as well as hadron properties based on quark-gluon degrees of freedom. Magnetic moments of the doubly-heavy baryons have been examined broadly in the literature, see \cite{Meng:2017dni,Li:2017cfz,Liu:2018euh,Blin:2018pmj,Ozdem:2018uue,Ozdem:2019zis} for recent references.

The magnetic moment of baryons can be obtained in terms of the spin, charge and effective mass of the bound quarks as
\begin{equation}
\mu_B= \sum_i \langle \phi_{sf} \vert \mu_{i}\vec{\sigma}^i \vert \phi_{sf} \rangle, \label{magnetic}
\end{equation}
where 
\begin{equation}
\mu_i=\frac{e_i}{2m_i^{eff}}.
\end{equation}
Here $e_i$ is the charge, $\sigma^i$ is the spin of the  constituent quark in the baryon, $\phi_{sf}$ is the spin-flavor wave function. The effective mass for the each constituent quark $m_i^{eff}$ can be defined as
\begin{equation}
m_i^{eff}=m_i \left( 1+ \frac{\langle H \rangle}{\sum_i m_i} \right),
\end{equation}
where $\langle H \rangle= E + \langle V_{S} \rangle$ and $m_i$'s are quark mass parameters of the respective model.

\subsection{Baryon Mass Inequality}
QCD predicts the existence of baryons containing not only a single heavy quark but also two or three heavy quarks ($c$ or $b$). Using necessary assumptions under some circumstances, hadrons satisfy mass inequalities associated with permutations of their quarks \cite{Bertlmann:1979zs,Richard:1983mu,Weingarten:1983uj,Witten:1983ut,Lieb:1985aw,Martin:1986da,Anwar:2017toa,Karliner:2019vhw}. 

By use of variational arguments and flavor independence, the following inequalities  can be written \cite{Nussinov:1983hb}:
\begin{eqnarray}
m_{x\bar{y}} &>& \frac{1}{2} (m_{x\bar{x}} + m_{y\bar{y}}) ~ \text{for ground state mesons} \\
m_{xyy} &>& \frac{1}{2} (m_{xxy} + m_{yyy}) ~ \text{for ground state baryons}.
\end{eqnarray}

For $ccu$ baryon, the following inequalities can be obtained in terms of $J^P$ notation \cite{Nussinov:1999sx}:
\begin{eqnarray}
m_{\Xi}(ccu) & \geq & (1/4) (2m_{J/\psi}+3 m_D + m_{D^*})~  \text{for} ~  (\frac{1}{2})^+, \label{ineq1} \\ 
m_{\Xi^\ast}(ccu) & \geq & (1/2) (m_{J/\psi} + 2m_{D^*}) ~ \text{for} ~  (\frac{3}{2})^+. \label{ineq2}
\end{eqnarray}
These inequalities enable rough estimates about some physical properties (mostly mass)  of unobserved hadron masses independent of potential models.

\section{\label{sec:level3}Numerical Results and Discussion}
The nonrelativistic Schrödinger equation for the model in this paper is solved numerically. Cornell potential parameters are taken from \cite{Debastiani:2017msn} whereas $u$ quark mass is taken from \cite{Griffiths:2008zz}. The parameters of the potential model are listed in Table \ref{tab:table1}.
\begin{table}[H]
\caption{\label{tab:table1}Quark model parameters of Cornell potential. }
\begin{ruledtabular}
\begin{tabular}{cc}
Parameter&Value \\
\hline
$m_u$& 0.336 ~ \text{GeV}  \\
$m_c$ & 1.4622 ~ \text{GeV} \\
$\sigma$ & 1.0831 ~ \text{GeV} \\
$b$ & 0.1463 ~ $\text{GeV}^2$  \\
$\alpha_s$ & 0.5202\\
\end{tabular}
\end{ruledtabular}
\end{table} 

\subsection{Mass Spectroscopy of $\Xi_{cc}^{++}$ in Quark-Diquark Model}
Before moving to the quark-diquark model calculations, an important warning should be made about the diquark masses. As mentioned before, diquark masses need to be calculated at the first step before delving into predictions of quark-diquark model. 

Due to the the SU(3) color symmetry of QCD, when we combine a quark and an antiquark in the fundamental color representation, we obtain $\vert q \bar{q} \rangle:  \textbf{3}  \bigotimes  \bar{\textbf{3}}= \textbf{1} \bigoplus \textbf{8} $. This representation gives the color factor for the color singlet as $\kappa=-4/3$ of the quark-antiquark system. In the diquark case, when we combine two quarks in the fundamental color representation, it reduces to $\vert q q \rangle:  \textbf{3}  \bigotimes  \textbf{3}= \bar{\textbf{3}} \bigoplus \textbf{6} $, a color antitriplet  $\bar{\textbf{3}}$ and a color sextet $\textbf{6}$.  The antitriplet state has a color factor $\kappa=-2/3$ which is attractive, whereas the sextet state  has a color factor $\kappa=+1/3$ which is repulsive. Changing  color factor $\kappa=-4/3$ of a quark-antiquark system to the color factor $\kappa=-2/3$ of a quark-quark system is equivalent of introducing a factor of 1/2 in the Coulomblike part of the Cornell potential for the quark-antiquark system. Furthermore, introducing a factor 1/2 could be taken as a global factor since it comes from the color structure of the wavefunction. Therefore the potential for diquark can be taken as $V_{qq}=V_{q\bar{q}}/2$ which implies $b \to b/2$ and $\kappa \to \kappa/2$ when dealing with diquarks. 

The quark content of $\Xi_{cc}^{++}$ is $ccu$. In this system, diquark clustering can occur between $u$ quark and $c$ quark as 
\begin{eqnarray}
\Xi_{cc}^{++} &=& \left[uc\right]c, \\
\Xi_{cc}^{++} &=& \left\lbrace uc \right\rbrace c,\\
\Xi_{cc}^{++} &=& \left\lbrace cc \right\rbrace u.
\end{eqnarray}
Here, $\left[ ab \right]$ indicates a $J^P=0^+$ diquark which is antisymmetric under  $a \leftrightarrow b$ and $\left\lbrace ab \right\rbrace$ indicates a $J^P=1^+$ diquark which is symmetric under $a \leftrightarrow b$ \cite{Yin:2019bxe}. The calculated diquark mass values are shown in Table \ref{tab:table2}. 

\begin{table}[H]
\caption{\label{tab:table2}Ground state $ccu$ mass values of different quark-diquark configurations. Mass values are in GeV.}
\begin{ruledtabular}
\begin{tabular}{ccccc}
State & Spin & Mass & Ref. \cite{Giannuzzi:2009gh} &Ref. \cite{Yin:2019bxe} \\
\hline
$\left[uc\right]$ & $S=0$& 2.150 & &2.15\\
$\left\lbrace uc \right\rbrace$ &$S=1$ &2.203&& 2.24\\
$\left\lbrace cc \right\rbrace$ &$S=1$ &3.133&3.238& 3.30 \\
\end{tabular}
\end{ruledtabular}
\end{table} 
The spin of a diquark can be one or zero. A diquark with two identical quark flavors can appear only in spin one state while a diquark with two different quark flavors can appear in both spin one and spin zero. The $cc$ color antitriplet diquark has spin one in order to make the wave function of the two quarks antisymmetric \cite{Jaffe:2004ph}. The spin of the third quark is either parallel to the diquark which gives total spin quantum number as $J=3/2$, or antiparallel to the diquark which makes total spin quantum number $J=1/2$. The mass values of $ccu$ baryon in different diquark configurations are given in Table \ref{tab:table3}.
\begin{table}[H]
\caption{\label{tab:table3}Ground state $ccu$ mass values of different quark-diquark configurations. Mass values are in GeV. }
\begin{ruledtabular}
\begin{tabular}{ccccc}
State & $J^P=\frac{1}{2}^+$ & $J^P=\frac{3}{2}^+$  \\
\hline
$\left[uc\right]_{S=0}c_{S=\frac{1}{2}}$ & 3.687& - \\
$\left\lbrace uc \right\rbrace_{S=1} c_{S=\frac{1}{2}}$ & 3.647 & 3.773\\
$\left\lbrace cc \right\rbrace_{S=1} u_{S=\frac{1}{2}}$ & 3.621 & 4.019\\
\end{tabular}
\end{ruledtabular}
\end{table} 
In literature, the $J=3/2$ state has been predicted to be heavier than the $J=1/2$ state by around 100 MeV. An interesting pattern resulted from $\left\lbrace cc \right\rbrace u$  structure is that, the hyperfine splitting is around 400 MeV although the mass of $J=1/2$ state agrees perfect with the experimental mass of $\Xi_{cc}^{++}$. At this point, it will be good to obtain the wave function at the origin. The value of the square modulus of the wave function at the origin $\vert \Psi(0) \vert^2$, is an important quantity. For example, decay widths can be calculated by using the wave function or its derivative at the origin. Wave function at the origin has non-zero value only for $S-$ wave ($\ell =0$) states. Therefore we have 
\begin{equation}
\vert \Psi(0) \vert^2= \vert Y^0_0(\theta, \phi) R_{n,\ell}(0)\vert^2=\frac{\vert R_{n,\ell}(0)\vert^2}{4\pi}.
\end{equation}
Here $\vert R_{n,\ell}(0)\vert^2$ can be obtained directly from the numerical calculations. We use the following formula, which relates the wave function at the origin $\vert \Psi(0) \vert^2$ to the radial potential \cite{Lucha:1991vn}:
\begin{equation}
\vert \Psi(0) \vert^2=\frac{\mu}{2 \pi} \langle \frac{d}{dr} V(r) \rangle \Rightarrow \vert R(0) \vert^2=2 \mu \langle \frac{d}{dr} V(r) \rangle.
\end{equation}
A large value of the wave function at the origin corresponds to a compact state. Wave functions at the origin $\vert R(0) \vert^2$ are calculated and the obtained results are given in Table \ref{tab:table4}.  
\begin{table}[H]
\caption{\label{tab:table4}Wave function at the origin of $ccu$ states in quark-diquark model. All results are in GeV$^3$. }
\centering
\begin{ruledtabular}
\begin{tabular}{cc}
State & $\vert R(0) \vert^2$  \\ \\
\hline
$\left[uc\right]_{S=0}c_{S=\frac{1}{2}}$ & 1.835 \\
$\left\lbrace uc \right\rbrace_{S=1} c_{S=\frac{1}{2}}$ & 1.875 \\
$\left\lbrace cc \right\rbrace_{S=1} u_{S=\frac{1}{2}}$ & 0.247 \\
\end{tabular}
\end{ruledtabular}
\end{table}
It can be seen from the table that $cc$ diquark clustering in $ccu$ state has significantly lower value of wave function at the origin than $cu$ diquark  in $ccu$ state. In \cite{Yin:2019bxe}, as a result of Faddeev amplitude it was found that dominant spin-flavor correlation to $J^P=\frac{1}{2}^+$ $\Xi_{cc}$ state comes from $\left[uc\right]c$ configuration whereas dominant spin-flavor correlation to $J^P=\frac{3}{2}^+$ $\Xi_{cc}^\ast$ state comes from $\left\lbrace uc \right\rbrace c$ configuration. Our results of the wave function at the origin support these outcomes of  \cite{Yin:2019bxe}. Although wave function at the origin depends significantly on the used potential and numerical method, similar result was obtained in \cite{SilvestreBrac:1996wp} where a Faddeev treatment had been used. The obtained results are 65.9 $\text{fm}^{-3}$ for $\Xi_{cc}(ucc)$ which is supposed to be composed of $cc$ pair and 119 $\text{fm}^{-3}$ for $\Xi_{cc}(ucc)$ which is supposed to be composed of $uc$ pair.  

In Table \ref{tab:table5} we present quark-diquark model predictions comparing with available results from the literature. Our model predicts compatible results compared to the references given in the related table. It can be seen from Table \ref{tab:table5} that our results of $\left[uc \right]c$, $\left\lbrace uc \right\rbrace c$ and $\left\lbrace cc \right\rbrace u$ states with $J^P=\frac{1}{2}^+$ are in good agreement with the reference studies. The mass of $\left\lbrace cc \right\rbrace u$ with $J^P=\frac{3}{2}^+$ is approximately 300 MeV higher than the corresponding $J^P=\frac{3}{2}^+$ states of references.

\begin{table*}[h]
\caption{\label{tab:table5}Comparison of ground state masses of $\Xi_{cc}^{++}$ baryon in quark-diquark model. In Ref. \cite{Yu:2018com}, [a] refers $m_Q \to \infty$, $\kappa=0.02$, [b] refers $m_Q \to \infty$, $\kappa=0.08$, [c] refers $m_Q$ finite, $\kappa=0.02$ and [d] refers $m_Q$ finite, $\kappa=0.08$. For more details, see the reference paper.}
\begin{ruledtabular}
\begin{tabular}{cccc}
References&$J^P=(\frac{1}{2})^+$&$J^P=(\frac{3}{2})^+$& Method\\
\hline
$\left[uc \right]c$  & 3.687& -& Quark-Diquark Model\\
$\left\lbrace uc \right\rbrace c$ & 3.647 & 3.773& \\
$\left\lbrace cc \right\rbrace u$ & 3.621 & 4.019 &\\
Ref. \cite{Roncaglia:1995az} &3.66 &3.74 &Feynman-Hellmann + Semiemperical\\
Ref. \cite{Ebert:1996ec} &  3.66& 3.81 & Relativistic Quark Model \\
Ref. \cite{Gershtein:2000nx} &3.478 &3.610 & Nonrelativistic Quark Model\\ 
Ref. \cite{Ebert:2002ig}&3.620 & 3.727 & Relativistic Quark Model\\
Ref. \cite{Roberts:2007ni} & 3.676& 3.753& Nonrelativistic Quark Model \\ 
Refs. \cite{Aliev:2012ru,Aliev:2012iv} & 3.72(0.20) & 3.69-3.72  & QCD Sum Rule \\
Ref. \cite{Shah:2017liu} &3.511 & 3.687& Hypercentral Constituent Quark Model\\
Ref. \cite{Vijande:2004at} & 3.524 & 3.548 & Nonrelativistic Quark Model + Potential Model\\
Ref. \cite{Giannuzzi:2009gh} & 3.547& 3719 & Bethe-Salpeter Model\\  
Refs. \cite{Wang:2010hs,Wang:2010vn} & 3.570&3.610 & QCD Sum Rule\\
Ref. \cite{Bali:2015lka} & 3.610(09)(12) & 3.694(07)(11)& Lattice QCD\\
Ref. \cite{Yoshida:2015tia} & 3.685 & 3.754 & Nonrelativistic Quark Model\\
Refs. \cite{Wei:2015gsa,Wei:2016jyk} & 3.520 & 3.695 & Regge Phenomenology\\
Ref. \cite{Alexandrou:2014sha} & 3.561(22) & 3.642(26)& Lattice QCD\\
Ref. \cite{Karliner:2014gca} &3.627 & 3.690 & Nonrelativistic Quark Model \\
Ref. \cite{Brown:2014ena} & 3.610 & 3.692 & Latice QCD\\
Ref. \cite{Lu:2017meb} &3.606 &3.675 & Relativistic Quark Model\\
Ref. \cite{Weng:2018mmf}  & 3.633& 3.696 & Chromagnetic Model\\
Ref. \cite{Yu:2018com}[a] &3.63$\pm$ 0.02 &3.63$\pm$ 0.02 & Bethe-Salpeter Model\\
Ref. \cite{Yu:2018com}[b]&3.62$\pm$ 0.02 &3.62$\pm$ 0.02 &\\
Ref. \cite{Yu:2018com}[c] &3.55$\pm$ 0.01 &3.62$\pm$ 0.01& \\
Ref. \cite{Yu:2018com}[d] &3.54$\pm$ 0.01 &3.62$\pm$ 0.01& \\
Ref. \cite{Wang:2018lhz} & $3.63^{+0.08}_{-0.07}$ & $3.75^{+0.07}_{-0.07}$ & QCD Sum Rule \\ 
Ref. \cite{Rahmani:2020pol} & 3.396 & 3.434 & Nonrelativistic Quark Model + Potential Model \\
Ref.  \cite{Li:2019ekr} & 3.601$^{-28}_{+28}$ &3.703$^{-28}_{+28}$ & Bethe-Salpeter Model\\  
Ref. \cite{Majethiya:2008ia} & 3.519 & 3.555 & Quark-Diquark Model\\
Ref. \cite{Zhang:2008rt}& 4.26 $\pm$ 0.19 &3.90 $\pm$ 0.10 & QCD Sum Rule \\
Ref. \cite{He:2004px} & 3.55 &3.59 & Bag Model \\ 
\end{tabular}
\end{ruledtabular}
\end{table*}
Our model for the $cc$ diquark clustering in ground state $ccu$ baryon gave a mass of 3621 MeV with $J^P=\frac{1}{2}^+$ which exactly matches the observed mass of $\Xi_{cc}^{++}$ baryon which is $3621.40 \pm 0.72(\text{stat.}) \pm 0.27 (\text{syst.})$ MeV. Ground state $J^P=\frac{3}{2}^+$ mass is approximately 400 MeV higher than the experimental mass of $\Xi_{cc}^{++}$. If $\Xi_{cc}^{++}$ has a $cc$ diquark clustering, its quantum numbers should be $J^P=\frac{1}{2}^+$. $\left[uc \right]c$ ground state with $J^P=\frac{1}{2}^+$ has a mass of 3687 MeV which is 66 MeV higher than the experimental mass of $\Xi_{cc}^{++}$. The mass of $\left\lbrace uc \right\rbrace c$  ground state with $J^P=\frac{1}{2}^+$ is 3647 MeV and 26 MeV above the experimental mass while the mass of $J^P=\frac{3}{2}^+$ is 3773 and 152 MeV above than the experimental mass. The mass spectrum analysis supports $J^P=\frac{1}{2}^+$ assignment for the $\Xi_{cc}^{++}$ baryon in all three diquark clustering cases. 

We also calculated the masses of radially excited states of $\Xi_{cc}^{++}$. The   results can be seen in Table \ref{tab:table6}. Our results are in good agreement with the available results of the Refs. \cite{Giannuzzi:2009gh} and \cite{Li:2019ekr}, in which both of the references used Bethe-Salpeter method. Analyzing the results,  it can be seen that mass results of 2S and 3S of $J^P=\frac{1}{2}^+$ $\left[uc \right]c$ states are in good agreement with the results of the Refs. \cite{Giannuzzi:2009gh} and \cite{Li:2019ekr}. This statement is valid for the results of 2S and 3S of $J^P=\frac{1}{2}^+$ $\left\lbrace uc \right\rbrace c$  states and 2S  and 3S of $J^P=\frac{3}{2}^+$ $\left\lbrace uc \right\rbrace c$ states. In the case of $\left\lbrace cc \right\rbrace u$, both $J^P=\frac{1}{2}^+$ and $J^P=\frac{3}{2}^+$ masses of radially excited states are considerably higher than the reference studies. 

\begin{table*}[h]
\caption{\label{tab:table6}Mass spectra of radial excited states of $\Xi_{cc}^{++}$ baryon in quark-diquark model.}
\centering
\begin{ruledtabular}
\begin{tabular}{cccccccccc}
State&$J^P$&This work&  \cite{Ebert:2002ig}&\cite{Roberts:2007ni}&\cite{Valcarce:2008dr} & \cite{Shah:2017liu} & \cite{Giannuzzi:2009gh}&\cite{Yoshida:2015tia} & \cite{Li:2019ekr} \\
\hline
&&$\left[uc \right]c$& && &  && \\
\hline
2S & $\frac{1}{2}^+$ & 4.274 & 3.910 & 4.029 & 3.976 & 3.920 & 4.183 & 4.079 &4.122$^{-38}_{+38}$ \\
3S &                 & 4.666 & 4.154 &       &       & 4.159 & 4.640 & 4.206 &4.504$^{-54}_{+54}$ \\
4S &                 & 4.992 &       &       &       & 4.501 &       &   \\
5S &                 & 5.281 &       &       &       & 4.748 &        &   \\
\hline
\hline
&&$\left\lbrace uc \right\rbrace c$& && &  && \\
\hline
2S & $\frac{1}{2}^+$ & 4.244 & 3.910 & 4.029 & 3.976 & 3.920 & 4.183 & 4.079 &4.122$^{-38}_{+38}$ \\
3S &                 & 4.646 & 4.154 &       &       & 4.159 & 4.640 & 4.206 &4.504$^{-54}_{+54}$ \\
4S &                 & 4.976 &       &       &       & 4.501 &       &   \\
5S &                 & 5.268 &       &       &       & 4.748 &        &   \\
\hline
2S & $\frac{3}{2}^+$ & 4.339 & 4.027 & 4.042 & 4.025 & 3.983 & 4.282 & 4.114 &4.232$^{-40}_{+40}$ \\
3S &                 & 4.725 &       &       &       & 4.261 & 4.719 & 4.131  \\
4S &                 & 5.048 &       &       &       & 4.519 &       &   \\
5S &                 & 5.334 &       &       &       & 4.759 &       &   \\
\hline
\hline
&&$\left\lbrace cc \right\rbrace u$& && &  && \\
\hline
2S & $\frac{1}{2}^+$ & 4.478 & 3.910 & 4.029 & 3.976 & 3.920 & 4.183 & 4.079 &4.122$^{-38}_{+38}$  \\
3S &                 & 5.026 & 4.154 &       &       & 4.159 & 4.640 & 4.206&4.504$^{-54}_{+54}$ \\
4S &                 & 5.482 &       &       &       & 4.501 &       &    \\
5S &                 & 5.888 &       &       &       & 4.748 &        &   \\
\hline
2S & $\frac{3}{2}^+$ & 4.670 & 4.027 & 4.042 & 4.025 & 3.983 & 4.282 & 4.114 &4.232$^{-40}_{+40}$ \\
3S &                 & 5.170 &       &       &       & 4.261 & 4.719 & 4.131  \\
4S &                 & 5.602 &       &       &       & 4.519 &       &   \\
5S &                 & 5.993 &       &       &       & 4.759 &       &   \\

\end{tabular}
\end{ruledtabular}
\end{table*}

\subsection{Magnetic Moment of $\Xi_{cc}^{++}$ in Quark-Diquark Model}
The explicit form of the spin-flavor wave functions of the spin-$\frac{1}{2}$ and spin-$\frac{3}{2}$ doubly-charmed $ccu$ baryons  can be written as:
\begin{eqnarray}
\vert \Xi_{ccu};s=\frac{1}{2}\rangle&=&\frac{1}{3\sqrt{2}}[2c\uparrow
c\uparrow u\downarrow-c\uparrow c\downarrow u\uparrow-c\downarrow
c\uparrow u\uparrow \nonumber \\ &+& 2c\uparrow u\downarrow c\uparrow-c\downarrow
u\uparrow c\uparrow -c\downarrow u\downarrow
u\downarrow \nonumber\\ &+& 2u\downarrow c\uparrow c\uparrow-u\downarrow c\downarrow
c\downarrow-u\uparrow c\downarrow c\uparrow],\\
\vert \Xi_{ccu}^{*};s=\frac{3}{2}\rangle&=&\frac{1}{\sqrt{3}}[
c\uparrow c\uparrow u\uparrow+c\uparrow c\uparrow
u\uparrow \nonumber \\ &+& c\uparrow c\uparrow u\uparrow],
\end{eqnarray}
where the arrows denote the third-components of the spin. The corresponding magnetic moments read as follows
\begin{eqnarray}
\vert \Xi_{cc}^{++}\rangle &=& \frac{4}{3}\mu_c - \frac{1}{3}\mu_u ~ \text{for} ~  J^P=\frac{1}{2}^+, \nonumber \\
\vert \Xi_{cc}^{++*}\rangle &=& 2\mu_c + \mu_u ~ \text{for} ~  J^P=\frac{3}{2}^+.
\end{eqnarray}
The obtained magnetic moments for $J^P=\frac{1}{2}^+$ $\Xi_{cc}^{++}$ and $J^P=\frac{3}{2}^+$ $\Xi_{cc}^{++*}$  are presented in Tables \ref{tab:table7} and \ref{tab:table8} with comparison of the available results in literature.

\begin{table*}[h]
\caption{\label{tab:table7}Comparison of the magnetic moments of  $J^P=\frac{1}{2}^+$ $\Xi_{cc}^{++}$ baryon with those predicted by other approaches (in unit of $\mu_N$).}
\begin{ruledtabular}
\centering
\begin{tabular}{ccc}
References&Magnetic Moment& Method\\
\hline
$\left[uc \right]c$  & -0.044 & Quark-Diquark Model \\
$\left\lbrace uc \right\rbrace c$ &-0.044 & \\
$\left\lbrace cc \right\rbrace u$ &-0.045 & \\
Ref. \cite{JuliaDiaz:2004vh} & -0.10 & Relativistic Quark Model \\
Ref. \cite{Faessler:2006ft} & 0.13 &Relativistic Three-Quark Model \\
Ref. \cite{Albertus:2006ya} & -0.208& Nonrelativistic Quark Model\\
Ref. \cite{Shah:2017liu} &0.031 & Hypercentral Constituent Quark Model\\
Ref. \cite{Ozdem:2018uue} & -0.23 $\pm$ 0.05 & QCD Sum Rule \\
Ref. \cite{Li:2020uok}&0.35 & Heavy Baryon Chiral Perturbation Theory \\
Ref. \cite{Li:2017cfz} & -0.25 & Heavy Baryon Chiral Perturbation Theory\\
Ref. \cite{Majethiya:2008ia} &-0.054 & Quark-Diquark Model\\
Ref. \cite{Patel:2007gx} & -0.133 & Hypercentral Constituent Quark Model\\  
Ref. \cite{Bernotas:2012nz} & 0.114 & MIT Bag Model\\
Ref. \cite{Lichtenberg:1976fi} & -0.12 & Quark Model \\
Ref. \cite{Oh:1991ws} & -0.47& Skyrmion (Set 1) \\
Ref. \cite{Oh:1991ws} & -0.47& Skyrmion (Set 2) \\
Ref. \cite{Sharma:2010vv} & 0.006& Chiral Constituent Quark Model\\
Ref. \cite{SilvestreBrac:1996bg}&-0.20 & Nonrelativistic Quark Model\\
Ref. \cite{Bose:1980vy} & 0.17 &MIT Bag Model \\
Ref. \cite{Jena:1986xs} &-0.154 & Logarithmic Confining Potential (Set 1) \\
Ref. \cite{Jena:1986xs} &-0.172 & Logarithmic Confining Potential (Set 2) \\
Ref. \cite{Simonis:2018rld} &-0.11 & MIT Bag Model\\
\end{tabular}
\end{ruledtabular}
\end{table*}
There is no data for the magnetic moments of doubly-heavy baryons including $\Xi_{cc}^{++}$. As can be seen from Table \ref{tab:table7}, our results of $\left[uc \right]c$,$\left\lbrace uc \right\rbrace c$ and $\left\lbrace cc \right\rbrace u$ are in good agreement with the result of  Ref. \cite{Majethiya:2008ia} in which they studied spectra and magnetic moments of ($ccq$ $q \in u,d,s$) systems using Coulomb plus Martin potential in quark-diquark model. The magnetic moments of $J^P=\frac{1}{2}^+$ $\Xi_{cc}^{++}$ change from -0.044 $\mu_N$ to 0.35 $\mu_N$. The results vary at most  0.40 $\mu_N$. 

\begin{table*}[h]
\caption{\label{tab:table8}Comparison of the magnetic moments of  $J^P=\frac{3}{2}^+$ $\Xi_{cc}^{++*}$ baryon with those predicted by other approaches (in unit of $\mu_N$).}
\begin{ruledtabular}
\centering
\begin{tabular}{ccc}
References&Magnetic Moment& Method\\
\hline
$\left\lbrace uc \right\rbrace c$ &2.347 &Quark-Diquark Model \\
$\left\lbrace cc \right\rbrace u$ &2.203 & \\
Ref. \cite{Albertus:2006ya} & 2.67& Nonrelativistic Quark Model\\
Ref. \cite{Shah:2017liu} &2.218 & Hypercentral Constituent Quark Model\\
Ref. \cite{Shi:2021kmm} & 3.50 & Covariant Baryon Chiral Perturbation Theory (EOMS) Case 1 \\ 
Ref. \cite{Shi:2021kmm} &3.51 &Covariant Baryon Chiral Perturbation Theory (HB) Case 1 \\ 
Ref. \cite{Shi:2021kmm} & 2.89 & Covariant Baryon Chiral Perturbation Theory (EOMS) Case 2 \\ 
Ref. \cite{Shi:2021kmm} &2.80 &Covariant Baryon Chiral Perturbation Theory (HB) Case 2 \\ 
Ref. \cite{Meng:2017dni} & 3.51 &Heavy Baryon Chiral Perturbation Theory (Set 1) \\  
Ref. \cite{Meng:2017dni} &3.63  & Heavy Baryon Chiral Perturbation Theory (Set 2) \\
Ref. \cite{Ozdem:2019zis} &2.94 $\pm$ 0.95 & QCD Sum Rule \\
Ref. \cite{Majethiya:2008ia} &2.513 & Quark-Diquark Model\\
Ref. \cite{Patel:2007gx} & 2.75 & Hypercentral Constituent 
Quark Model\\ 
Ref. \cite{Oh:1991ws} & 3.16& Skyrmion (Set 1) \\
Ref. \cite{Oh:1991ws} & 3.18& Skyrmion (Set 2) \\
Ref. \cite{Sharma:2010vv} & 2.66& Chiral Constituent Quark Model\\
Ref. \cite{Bose:1980vy} & 2.54 &MIT Bag Model \\
Ref. \cite{Simonis:2018rld} &2.35 & MIT Bag Model\\
Ref. \cite{Dhir:2009ax} &2.41 &Effective Mass Scheme \\
Ref. \cite{Dhir:2009ax} &2.52 &Screening Effect Scheme \\
\end{tabular}
\end{ruledtabular}
\end{table*}

In the case of $J^P=\frac{3}{2}^+$ $\Xi_{cc}^{++*}$ baryon, as can be seen from Table \ref{tab:table8}, our result of $\left\lbrace uc \right\rbrace c$ baryon is in good agreement with the MIT Bag Model of Ref. \cite{Simonis:2018rld} and effective mass scheme of Ref. \cite{Dhir:2009ax}, while our result of $\left\lbrace cc \right\rbrace u$ baryon is in good agreement with the hypercentral constituent quark model of  Ref. \cite{Shah:2017liu}. The magnetic moments of $J^P=\frac{3}{2}^+$ $\Xi_{cc}^{++}$ change from 2.203 $\mu_N$ to 3.51 $\mu_N$. The results vary at most  1.3 $\mu_N$. 

According to Tables \ref{tab:table7} and \ref{tab:table8}, it can be said that the theoretical results are very much scattered. More experimental investigations are needed to understand the current situation. Such studies may provide vital information on the nature of $\Xi_{cc}^{++}$ baryon. 

\subsection{Baryon-Mass Inequality of $\Xi_{cc}^{++}$ in Quark-Diquark Model}
In this part, we have compared our results of quark-diquark model for the baryon mass inequality. For this reason, we have used Particle Data Group data for the masses of $J/\psi$, $D$ and $D^\ast$ mesons \cite{Zyla:2020zbs}. The results are presented in Table \ref{tab:table10}. The mass predictions of doubly-charmed heavy baryons are found to be well above the equality limit, satisfying the inequality relations, given in Eqs. (\ref{ineq1}) and (\ref{ineq2}).

\begin{table}[h]
\caption{\label{tab:table10} Baryon mass inequalities of the  Quark-Diquark Model.}
\begin{ruledtabular}
\centering
\begin{tabular}{cccccc}
Model & $J^P=\frac{1}{2}^+$ & Eq. \ref{ineq1} & $J^P=\frac{3}{2}^+$ & Eq. \ref{ineq2}\\ 
\hline
$\left[uc\right]c$ & 3.687 & 3.448 &- & 3.555 \\
$\left\lbrace uc \right\rbrace c$ & 3.647 & 3.448 & 3.773 & 3.555 \\
$\left\lbrace cc \right\rbrace u$ & 3.621 & 3.448 & 4.019& 3.555 \\ 
\end{tabular}
\end{ruledtabular}
\end{table}

\section{\label{sec:level4}Concluding Remarks and Final Notes}
The observation of $\Xi_{cc}^{++}$ has aroused tremendous attention to the doubly-charmed baryon systems. In this present work, we studied mass spectra and magnetic moments of doubly-charmed baryons by using a potential model in which two quarks inside the baryon are assumed to be clustered into a diquark configuration. In this model, the interaction between two colored particles is taken to be a Coulomb-like potential at short distances transforming like a component of a four-vector and a linear potential at large distances transforming like a Lorentz scalar. The color factor $\kappa$ is -2/3 for two quarks in a diquark and  -4/3 for a quark and diquark inside a baryon. The parameters of this model are obtained by fitting the meson spectrum. 

The diquark correlations are essentially dynamical and play a key role in formulating and solving the three valence-quark baryon problems. In a quark-diquark model, we have the problem of deciding which two quarks are bound in the diquark. For doubly-heavy baryons, it is natural to think that two heavy quarks cluster in the baryon. We have taken into account all the possible diquark clusterings inside the $ccu$ baryon. These are antisymmetric and symmetric diquark clustering of $uc$ and symmetric clustering of $cc$, i.e. spin-0 and spin-1 diquark cores. The obtained ground state mass results of $ccu$ baryon in different diquark clustering agree well in the case of $J^P=\frac{1}{2}^+$ $\left[uc\right]c$, $\left\lbrace uc \right\rbrace c$ and $\left\lbrace cc \right\rbrace u$. The results of $J^P=\frac{3}{2}^+$ $ccu$ baryon agree well in the $\left\lbrace uc \right\rbrace c$ state but it is approximately 400 MeV above for the $\left\lbrace cc \right\rbrace u$ state with respect to reference studies. Masses of  radially excited states are significantly above than the reference studies. 

Wave function at the origin of the corresponding $\left[uc\right]c$,$\left\lbrace uc \right\rbrace c$ and $\left\lbrace cc \right\rbrace u$ states are calculated. Our results suggest that the $uc$ diquark clustering is more favourable than the $cc$ diquark clustering which is widely accepted point of view for $\Xi_{cc}^{++}$ baryon. 

We also calculated the magnetic moments of spin-1/2 and 3/2 of $\Xi_{cc}^{++}$ baryon. The magnetic moments of doubly-heavy baryons encode key knowledge about their internal structure and shape deformations. Comparing our results with the predicted magnetic moments of other theoretical models leads to rather different estimations for spin-1/2 and spin-3/2. Due to the lack of experimental and lattice QCD data, such efforts are important for the observed $\Xi_{cc}^{++}$ state, especially for the spin-1/2 case. The direct measurement of the magnetic moment of the spin-3/2 doubly-heavy baryons is unlikely in the near future. Therefore, any even non-straightforward approximation of the magnetic moments of doubly-heavy baryons may be helpful. 

We hope our numerical calculation for masses of the ground and radially excited states may be helpful for identifying the type of diquark clustering in $\Xi_{cc}^{++}$. We also hope that our results may be useful for future experimental measurement of magnetic moments of the doubly-charmed baryons.

\end{document}